\documentclass[preprint]{aastex}
\shorttitle{PECULIAR MOTION OF THE SOLAR SYSTEM FROM THE SKY BRIGHTNESS SKY BRIGHTNESS}
\shortauthors{SINGAL}
\begin{document}
\title{LARGE PECULIAR MOTION OF THE SOLAR SYSTEM FROM THE DIPOLE ANISOTROPY IN SKY BRIGHTNESS DUE TO DISTANT RADIO SOURCES}
\author{ASHOK K. SINGAL}
\affil{ASTRONOMY \& ASTROPHYSICS DIVISION, PHYSICAL RESEARCH LABORATORY, \\
NAVRANGPURA, AHMEDABAD - 380 009, INDIA; asingal@prl.res.in}

\begin{abstract} 
According to the cosmological principle, the Universe should appear isotropic, without any preferred directions, 
to an observer whom we may consider to be fixed in the co-moving co-ordinate system of the expanding Universe. 
Such an observer is stationary with respect to the average distribution of the matter in the Universe and the 
sky brightness at any frequency should appear uniform in all directions to such an observer. 
However a peculiar motion of such an observer, due to a combined effect of Doppler boosting and aberration,  
will introduce a dipole anisotropy in the observed sky brightness;
in reverse an observed dipole anisotropy in the sky brightness could be 
used to infer the peculiar velocity of the observer with respect to the average Universe. 
We determine the peculiar velocity of the solar system relative to the 
frame of distant radio sources, by studying the anisotropy in the sky brightness from discrete radio sources,
i.e., an integrated emission from discrete sources per unit solid angle.
Our results give a direction of the velocity vector in agreement with the Cosmic Microwave Background 
Radiation (CMBR) value, but the magnitude ($\sim 1600\pm 400$ km/s) is $\sim 4$ times the CMBR value ($369\pm 1$ km/s)
at a statistically significant ($\sim 3\sigma$) level. 
A genuine difference between the two dipoles would imply anisotropic 
Universe, with the anisotropy  changing with the epoch. This would violate the cosmological principle where 
the isotropy of the Universe is assumed for all epochs, and on which the whole modern cosmology is based upon.
\end{abstract}


\maketitle
\section{INTRODUCTION}
\begin{figure}[h]
\epsscale{.70}
\plotone{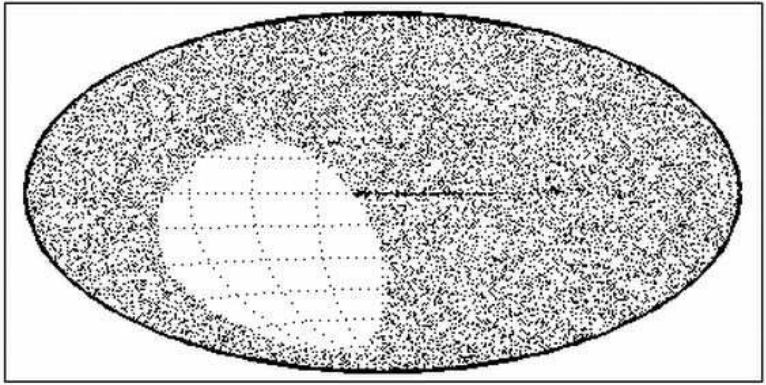}
%\scalebox{0.5}{\includegraphics{singalfig1.eps}}
\caption{The distribution of strong NVSS sources ($S>300$ mJy) in galactic co-ordinates}
\end{figure}

The peculiar velocity of the solar system through the Universe has been determined relative to the frame of 
reference provided by the Cosmic Microwave Background Radiation (CMBR), to be $369$ km/s in the direction 
$l=264^{\circ}, b=48^{\circ}$ (Lineweaver et al. 1996; Hinshaw et al. 2009). A study  using the number counts 
of radio sources (Blake and Wall 2002) had found the velocity to be consistent with that from the CMBR. 
Here we determine the peculiar motion from the anisotropy in sky brightness due to radio sources.
This provides an independent check on the interpretation of CMBR dipole anisotropy being due to motion of the 
solar system. Also CMBR provides information 
about the isotropy of the Universe for redshift $z \sim 700$, but the radio source population refers to a 
much later epoch $z \sim 1$. Thus it also provides an independent check on the cosmological principle 
where isotropy of the Universe is assumed for all epochs. 
\section{THE SOURCE CATALOGUE}
We have used the NVSS catalogue (NRAO VLA Sky Survey, Condon et al. 1998) for our investigations. This survey
covers whole sky north of declination $-40^{\circ}$, a total of 82\% of the celestial sphere, at 1.4 GHz. 
There are about 1.8 million sources in the catalogue with a flux-density limit $S>3$ mJy. 
Figure~1 shows a plot of the bright sources ($>300$ mJy) from the NVSS catalogue in galactic co-ordinates. 
There is a strip of excess sources 
near the galactic equator. The large gap corresponds to the southern declination limit of the survey.
\section{DIPOLE ANISOTROPY DUE TO DOPPLER BOOSTING AND ABERRATION}
An observer moving with a velocity $v$, will find sources in the forward direction brighter 
by a factor  $\delta^{1+\alpha}$, due to Doppler boosting, where $\delta=1+(v/c)\cos\theta$ is the 
Doppler factor, $c$ is the velocity of light and $\alpha$ ($\approx 0.8$) is the spectral index defined by 
$S \propto \nu^{-\alpha}$.
Here we have used the non-relativistic formula for the Doppler factor as CMBR observations indicate that $v\ll c$.
The increased flux density due to Doppler boosting in the forward direction will cause a telescope of a given 
sensitivity limit to detect comparatively a larger number of sources.
The contribution to the observed sky brightness at a given flux-density level $S$ comes from sources having the  
rest-frame flux-density $S/\delta^{1+\alpha}$. 
With the integral source counts of extragalactic radio source population following a power law  
$N(>S)\propto S^{-x}$ ($x \sim 1$, Ellis and Baldwin 1984), the number of the sources observed therefore 
will be higher by $\delta^{x(1+\alpha)}$, a factor independent of $S$ (as long as $x$ and $\alpha$ can be 
deemed to be independent of $S$ level). The observed sky 
brightness will thus change by an overall factor $\delta^{x(1+\alpha)}$ due to Doppler boosting. 

Also due to the aberration of light, the apparent position of a source at angle $\theta$ will be shifted 
in the forward direction by an angle $v \sin \theta/c$, 
as a result there will be a higher number density per steradian in the forward direction as compared to that 
in the backward direction. This excess in number density $\propto \delta^2$.
Thus as a combined effect of Doppler boosting and the aberration, the observed sky brightness 
(an integrated emission from discrete sources per unit solid angle) will vary as $\propto \delta^{2+x(1+\alpha)}$ 
which (for $v \ll c$) can be written as $1+{\cal D}\cos\theta$, a dipole anisotropy over the sky 
with amplitude ${\cal D}=[2+x(1+\alpha)](v/c)$ (Ellis and Baldwin 1984). 

To find the velocity of the solar system, we consider all sources to lie on the surface of a sphere of 
unit radius, and let $\textbf {r}_i$ be the position 
vector of $i^{th}$ source with respect to the centre of the sphere. An observer stationary at the centre of the 
sphere will find the sky brightness to be uniformly distributed in all directions (due to the assumed isotropy 
of the Universe) and therefore should get $\Sigma S_i \textbf{r}_i=0$. On the other hand for a moving observer, 
the forward shift in position due to aberration and the Doppler boosting of  
flux density in the forward direction, implies that the vectorial sum $\Sigma S_i\textbf{r}_i$ 
will yield a net vector in the direction of motion, thereby fixing the direction of the dipole.
If $\theta_i$ is the polar angle of the $i^{th}$ source with respect to the dipole direction, then 
the magnitude of the vectorial sum can be written as ${\Delta {\cal F}}=\Sigma S_i\: \cos \theta_i$. Writing 
${\cal F}=\Sigma S_i\: |\cos \theta_i|$ and converting the summation into an integration over the sphere, we get
\begin{equation}
\frac {\Delta {\cal F}}{{\cal F}}=k\: \frac {\int^{\pi}_{0}(1+{\cal D}\cos\theta)\cos \theta\: \sin \theta\:{\rm d}\theta}
{2\int^{\pi/2}_{0}\cos \theta \:\sin \theta\:{\rm d}\theta}=\frac {2k {\cal D}}{3}= \frac{2k}{3} \left[2+x(1+\alpha)\right]\frac{v}{c}.
\end{equation}
The formula is equally valid for samples with finite upper and lower flux-density limits. Here $k=1$ for a   
sky fully covered by the sample, and $1<k\leq 3/2$ when there are finite gaps in the sky coverage and may need to be 
determined numerically for individual cases. 

As the NVSS catalogue has a gap of sources for Dec $<-40^{\circ}$, in that case our assumption of 
$\Sigma S_i \textbf{r}_i=0$ for a stationary observer does not hold good. However if we drop all sources with  
Dec $> 40 ^{\circ}$ as well, then with equal and opposite gaps on opposite 
sides  $\Sigma S_i \textbf{r}_i=0$ is valid for a stationary observer. 
Further we also excluded all sources from our sample which lie in the galactic plane ($|b|<10^{\circ}$)
as the excess of galactic sources towards the galactic centre (see Fig.~1) is likely to contaminate the determination of 
velocity. Of course exclusion of such strips, which affect the forward and backward measurements 
identically, to a first order do not have systematic effects on the results (Ellis and Baldwin 1984). 
\begin{table}[h]
\begin{center}
\caption{The velocity vector from the dipole asymmetry in sky brightness}
\hskip4pc\vbox{\columnwidth=33pc
\begin{tabular}{cccccccccc}
\tableline\tableline 
 Flux-density Range & N && ${\cal D}$   & &  RA && Dec && $v$\\
 (mJy) && & ($10^{-2}$) && ($^{\circ}$)& & ($^{\circ}$) && ($10^{3}$ km/s) \\ \hline
$1000>S\geq 50$ & 090360 & &$2.3\pm 0.7$ && $163 \pm 12$ && $-11 \pm 11$ && $1.8\pm0.5$ \\
$1000>S\geq 40$ & 114600 & &$2.2\pm 0.6$ && $159 \pm 12$ && $-11 \pm 11$ && $1.7\pm0.5$ \\
$1000>S\geq 35$ & 131691 & &$2.2\pm 0.6$ && $159 \pm 11$ && $-10 \pm 10$ && $1.7\pm0.5$ \\
$1000>S\geq 30$ & 153759 && $2.2\pm 0.6$ && $159 \pm 11$ && $-07 \pm 10$ && $1.8\pm0.5$ \\
$1000>S\geq 25$ & 184237 && $2.2\pm 0.6$ & &$159 \pm 10$ && $-07 \pm 09$ && $1.7\pm0.4$ \\
$1000>S\geq 20$ & 228128 && $2.1\pm 0.5$ & &$158 \pm 10$ && $-06 \pm 09$ && $1.7\pm0.4$ \\
$1000>S\geq 15$ & 296811 && $2.0\pm 0.5$ & &$157 \pm 09$ && $-03 \pm 08$ && $1.6\pm0.4$ \\
\tableline
\end{tabular}
}
\end{center}
\end{table}

Before proceeding with the actual source sample we used the Monte--Carlo technique to create an artificial radio sky with about 
two million sources (number density of sources similar to that in the NVSS catalogue) distributed at random positions in 
the sky. While the sky positions for each simulation were allotted randomly for each source, for the flux-density 
distribution we took the observed NVSS sample, the latter justified because the source counts remain unchanged 
when integrated over the whole sky. The aberration merely shifts the apparent positions in the sky without adding or 
removing any sources and the 
effects of Doppler boosting on the differential source counts at any flux-density level are equal and opposite in the 
forward and backward hemispheres, therefore the total source counts summed over the whole sky remain the same. 
On this we superimposed Doppler boosting and aberration effects of our assumed motion, choosing a different velocity 
vector for each simulation (in a number of cases we also used the velocity vector from CMBR measurements 
(Lineweaver et al. 1996; Hinshaw et al. 2009) for Monte--Carlo simulations). 
The resultant artificial sky was then used to recover back the velocity vector under 
conditions similar to our NVSS case (e.g., with $|{\rm Dec}|> 40 ^{\circ}, |b|<10^{\circ}$ gaps in the sky), and thence 
obtained velocity vector was compared with the value actually used in that particular simulation. This not only verified our 
procedure but also allowed us to make an estimate of errors in the dipole co-ordinates as a large number of simulations 
($\sim 200$) were run starting with different random sky positions and for a different velocity vector each time. 
From these simulations we also estimated $k \sim 1.1$ (see Eq.~(1)), for the effect of the gaps in our samples. 
\section{RESULTS AND DISCUSSION}
\begin{table}[h]
\begin{center}
\caption{Speed determined from sky brightness for different $|{\rm sgb}|$ limits}
\hskip4pc\vbox{\columnwidth=33pc
\begin{tabular}{cccccccccc}
\tableline \tableline
 Flux-density Range & $|{\rm sgb}|\ge0^{\circ}$ && $|{\rm sgb}|\ge5^{\circ}$ && $|{\rm sgb}|\ge10^{\circ}$   & &  $|{\rm sgb}|\ge15^{\circ}$ && $|{\rm sgb}|\ge20^{\circ}$ \\
 (mJy)  & ($10^{3}$ km/s) &&($10^{3}$ km/s) &&($10^{3}$ km/s) & & ($10^{3}$ km/s)&&($10^{3}$ km/s) \\ \hline
$1000>S\geq 50$ & $1.8\pm0.5$ && $1.8\pm0.6$ && $1.8\pm0.6$&&  $2.0\pm0.6$ &&  $1.9\pm0.7$  \\
$1000>S\geq 40$ & $1.7\pm0.5$ && $1.7\pm0.5$ && $1.7\pm0.6$&&  $1.8\pm0.6$ &&  $1.8\pm0.6$  \\
$1000>S\geq 35$ & $1.7\pm0.5$ && $1.7\pm0.5$ && $1.7\pm0.5$&&  $1.8\pm0.6$ &&  $1.8\pm0.6$  \\
$1000>S\geq 30$ & $1.8\pm0.5$ && $1.7\pm0.5$ && $1.7\pm0.5$&&  $1.8\pm0.5$ &&  $1.8\pm0.6$  \\
$1000>S\geq 25$ & $1.7\pm0.4$ && $1.7\pm0.5$ && $1.6\pm0.5$&&  $1.7\pm0.5$ &&  $1.7\pm0.5$  \\
$1000>S\geq 20$ & $1.7\pm0.4$ && $1.6\pm0.4$ && $1.6\pm0.5$&&  $1.7\pm0.5$ &&  $1.7\pm0.5$  \\
$1000>S\geq 15$ & $1.6\pm0.4$ && $1.6\pm0.4$ && $1.6\pm0.4$&&  $1.6\pm0.5$ &&  $1.6\pm0.5$  \\
\tableline
\end{tabular}
} 
\end{center}
\end{table} 
Our results are presented in Table~1, which is almost self-explanatory. 
As a relatively small number of strong sources at high flux-density levels could introduce large statistical 
fluctuations in the sky brightness, we have restricted our sample to below 1000 mJy level. At the lower end 
we have restricted it to 15 mJy levels as the completeness of the sample at weaker flux-density levels could be 
doubtful (Blake and Wall 2002). 
The error in ${\cal D}$ due to statistical fluctuations comprises two components. 
The variance in the flux-density distribution about the average value $S_0$ among sources  
contributes an error $\sigma_1=(\Sigma (S_i - S_0)^2\cos^2 \theta_i )^{1/2}=(\Sigma S_i^2 - N S_0^2 )^{1/2}/{\sqrt 3}$, 
while the statistical fluctuations in number density 
in the two hemispheres contributes $\sigma_2=\sqrt N S_0/{\sqrt 3}$. As these two are statistically independent,  
the net error in ${\Delta {\cal F}}/{\cal F}$ then is
$(\sigma_1^2+\sigma_2^2)^{1/2}/{\cal F}=(\Sigma S_i^2)^{1/2}/{\sqrt 3}{\cal F}$.
The effects of error in the dipole direction on $\Delta F$ can be estimated this way. A shift $\Delta \theta$ in the dipole 
direction causes an exchange of spherical wedges of solid angle $2 \Delta \theta$ near the equator 
(i.e. for $\theta \sim \pi/2$) between the forward and backward hemi-spheres. However the dipole anisotropy 
being minimum there ($\propto \cos\theta$) the effect on speed estimate is minimal. It 
will result in a fractional change in ${\Delta {\cal F}}$ of about $4 \sin^2(\Delta \theta/2) (2\Delta \theta/2\pi)(3/2)
\sim 3 (\Delta \theta)^3 /2\pi$. In all cases it will be a systematic effect, resulting in
the speed being slightly underestimated ($\stackrel{<}{_{\sim}} 1 \%$ for a typical error of 12-15 deg in the dipole direction).

From Table~1 the direction the velocity vector (with our best estimate 
RA$=157^{\circ} \pm 9^{\circ}$, Dec$=-03^{\circ} \pm 8^{\circ}$ 
or in galactic co-ordinates $l=248^{\circ}\pm 12^{\circ}, b=44^{\circ}\pm 8^{\circ}$) is 
quite in agreement with those determined from the CMBR 
(RA$=168^{\circ}$, Dec$=-7^{\circ}$ or $l=264^{\circ}, b=48^{\circ}$ with errors less than a degree)
(Lineweaver et al. 1996; Hinshaw et al. 2009). However the estimates of $v$ 
($\sim 1.6\pm0.4 \times 10^3$ km/s) appear much higher than the CMBR value ($369\pm 1$ km/s) by a factor 
$\sim 4$ at a statistically significant ($\sim 3\sigma$) level. 

To guard against the possibility that some systematic effects like local clustering (mainly the Virgo 
super-cluster) might have affected the dipole magnitude, we restricted our region of the sky brightness to that outside the 
super-galactic plane by rejecting sources with low super-galactic latitude, $|{\rm sgb}|$. 
We determined the dipole progressively excluding sources in the latitude steps of 5 degrees, 
and from a comparison of all these cases ($|{\rm sgb}| \ge 0^{\circ}, 5^{\circ},10^{\circ},15^{\circ},20^{\circ}$) no systematic 
changes were seen in the computed dipole magnitude (Table~2). 
Thus it does not seem that the observed $v$, a factor of $\sim 4$ larger than the CMBR, has 
resulted from a local clustering. 
\begin{table}[h]
\begin{center}
\caption{The velocity vector from the number counts}
\hskip4pc\vbox{\columnwidth=33pc
\begin{tabular}{cccccccccc}
\tableline\tableline 
S & N && $\cal D$   & &  RA && Dec && $v$\\
 (mJy) && & ($10^{-2}$) && ($^{\circ}$)& & ($^{\circ}$) && ($10^{3}$ km/s) \\ \hline
$\geq 50$ & 091957 & & $2.1\pm 0.5$ &&  $171\pm 13$ &&  $-18\pm 14$ && $1.7\pm0.4$ \\
$\geq 40$ & 115837 & & $1.8\pm 0.4$ &&  $158\pm 12$ &&  $-19\pm 12$ && $1.4\pm0.4$ \\
$\geq 35$ & 132930& & $1.9\pm 0.4$ &&  $157\pm 11$ &&  $-12\pm 11$ && $1.5\pm0.3$ \\
$\geq 30$ & 154996 &&  $2.0\pm 0.4$ &&  $156\pm 11$ &&  $-02\pm 10$ && $1.6\pm0.3$ \\
$\geq 25$ & 185474 && $1.8\pm0.4$ & & $158\pm 11$ &&  $-02\pm 10$ && $1.4\pm0.3$ \\
$\geq 20$ & 229365 &&  $1.8\pm0.3$ & & $153\pm 10$ &&  $+02\pm 10$ && $1.4\pm0.3$ \\
$\geq 15$ & 298048 &&  $1.6\pm0.3$ & & $149\pm 09$ &&  $+15\pm 09$ && $1.3\pm0.2$ \\
\tableline
\end{tabular}
}
\end{center}
\end{table}
 
An earlier attempt using the number counts  of radio sources (Blake and Wall 2002) had found a peculiar velocity seemingly 
consistent with that from the CMBR observations, but from the sky brightness anisotropy we got the velocity $\sim 4$ times 
the CMBR value. 
To ascertain that the difference somehow is not between the dipoles arising from the sky brightness and the number counts, 
we have determined the velocity from the number counts as well, using a technique slightly different from that of 
Blake and Wall (2002). First the direction of the dipole was determined from $\Sigma\textbf{r}_i$, the three vector 
components ($x,y,z$) being essentially the same as those of the dipole position determined from the three $l=1$ 
spherical harmonic coefficients (Blake and Wall 2002).  Then the dipole magnitude was calculated from the fractional difference 
${\Delta {\cal N}}/{{\cal N}}= \Sigma \cos \theta_i\;/\;\Sigma |\cos \theta_i|=(2k/3) [2+x(1+\alpha)](v/c)=2k {\cal D}/3$, 
similar to that for ${\Delta {\cal F}}/{\cal F}$ in the case of sky brightness. The results are summarized in Table~3. 
Comparing with Table~1 we notice that the observed anisotropies in both the sky brightness and the number counts yield  
similar velocities with magnitudes $\sim 4$ times the CMBR value in both cases. 
However while in number counts the weaker 
sources, because of their much larger numbers ($\propto S^{-x}$), dominate the dipole determination, in the case of sky 
brightness, the contribution of each 
source being proportional to its flux density, the dipole determination depends equally on the stronger sources,  
$ S^{-x} \times S \sim 1$ (for $x\sim 1$).

A comparison of our results for source counts (Table~3) with those of Blake and Wall (2002)  
shows that the direction estimates of the dipole match exceedingly well, with an almost one to one correspondence for 
various bins. In fact, at a first look, even the magnitudes of dipoles, with average value ${\cal D}\sim 1.9\times 10^{-2}$ in 
our case as compared to $\sim 1.8\times 10^{-2}$ in Blake and Wall (2002), seem to match very well. However, there is an 
essential difference. The dipole magnitude defined by Blake and Wall (2002) as $2(2+x(1+\alpha))v/c$, 
is a factor of 2 larger than ${\cal D}$ defined in our case. Thus while the dipole expected from the CMBR value in 
Blake and Wall (2002) is $\sim 0.9\times 10^{-2}$, in our case it is only half of that ($\sim 0.47\times 10^{-2})$.
The tabulated values of  Blake and Wall (2002) seem about a factor of 2 larger 
than the CMBR prediction in all bins ($\sim 1.8\times 10^{-2}$ vs. $\sim 0.9\times 10^{-2}$)
though only at $\sim 1.5\sigma$ level.
But in our case the observed dipole is $\sim 4$ times the CMBR prediction ($\sim1.9\times 10^{-2}$ vs. 
$\sim 0.47\times 10^{-2}$) at $\sim 3\sigma$ level. 
Thus effectively we are finding dipole magnitude to be double of that by Blake and Wall (2002),  
which is quite surprising since the 
basic data used (NVSS) is the same  even if the techniques differ. In a consistent case our tabulated  
${\cal D}$ for all bins should have been $\sim 0.9\times 10^{-2}$ (i.e., half of the tabulated values of 
Blake and Wall 2002), which  definitely is not the case. 

The major difference in the two techniques is in the masking out of certain areas in sky which may otherwise 
contribute excess local sources to cause an under or overestimate of the dipole magnitude. However, it will be 
very surprising if this were making the difference of 2 for almost all flux bins. After all the local cluster sources 
should have different source counts than 
the truly far-off sources and should be affecting different flux bins quite differently; the difference should have been  
especially discernible if the overall source counts are going to get affected by a factor of $\sim 4$ or so.  It will be all 
the more intriguing if it had to happen without causing any shifts in the direction of the dipole which in both 
cases is found to be the same as that of the CMBR dipole.

To us it appears that more likely the actual dipole determination in both cases 
(Blake and Wall 2002 and ours) is essentially the same (with the different techniques 
and differential masking procedure making $\stackrel{<}{_{\sim}} 10 \%$ difference) and that the difference of 2 
creeps in while relating the dipole magnitude to the velocity as the two formulae differ by a factor of 2. 
We can only state that because of the discrepancy we are finding with the already known results 
(Lineweaver et al. 1996; Blake and Wall 2002; Hinshaw et al. 2009) we had to be extra careful in 
the magnitude scaling as well as in checking our code thoroughly, through Monte-Carlo simulations and otherwise.

One way to conclusively eliminate the possibility that our large dipole value could be a consequence of some unknown or 
ignored local clustering in certain regions of the sky is to determine the projection of the inferred velocity for 
different polar angles with respect to the dipole direction and to verify if the observed velocity components are  
following the $v \cos\theta$ relation. This is because any local clustering would affect the magnitude of the components 
being determined from different regions of the sky very differently. We determined the velocity components 
for three different polar angles for the last flux-density bin ($>15$ mJy) which had the largest number of sources. 
Starting from the dipole direction we first divided the sky into six equal-area zones. Then computing fractional 
difference in source counts between symmetrically placed pairs of sky zones, we determined the peculiar velocity components 
for three different polar angles. Figure~2 shows a plot of the  
the three components, which seem to fit very well with the expected $\cos \theta$ variation of $v=1300$ km/s, the 
estimated speed for that bin in Table~3. Also plotted is the projection of the CMBR value (369 km/s) for a comparison. 
\begin{figure}[h]
\epsscale{.70}\plotone{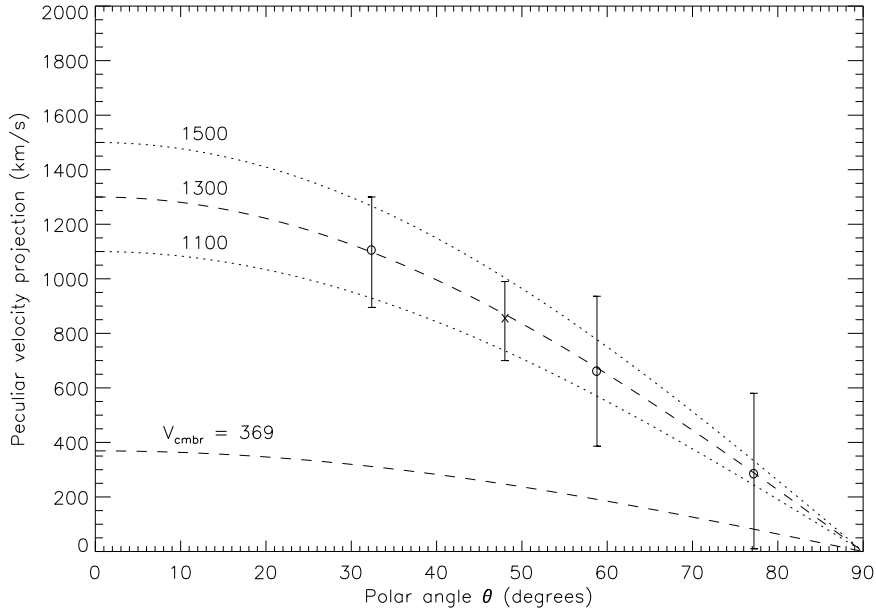}
%\scalebox{0.7}{\includegraphics{singalfig2.eps}}
\caption{A plot of the observed peculiar velocity component for different polar angles. Circles (o) 
represent values for velocity components obtained for three different angle from six equal area slices of the sky. Cross 
(x) marks the value obtained from the whole sky and which had provided the dipole magnitude for the corresponding 
($>15$ mJy) bin in Table ~3. The CMBR value ($V_{\rm cmbr}=369$ km/s) is shown for a comparison.
The dotted lines represent the $1\sigma$ error limits about expected component values for the 1300 km/s speed.}
\end{figure}

Suppose a relatively local (but perhaps on a scale much larger than that of Virgo supercluster but at $z\ll 1$) over-density 
generating the gravitational field responsible for the large observed motion, yields on an over-density also in the distribution 
of radio sources, and it gives a significant contribution to the 
excess sky brightness and to the excess counts in the direction of motion. 
The observed dipole amplitude will then be the sum of the contribution due to the actual motion with the added contribution of the 
source over-density masquerading as high solar motion in our analysis, in that case the derived velocity could be an upper limit. 
But Fig.~2 would be still difficult to explain there. 
If some excess in the sources due to local clustering in certain regions of the sky were indeed 
instrumental in causing the tabulated large ($v=1300$ km/s) value, then the three 
velocity component values could not have been influenced by the same factor almost 
identically by the local clustering as these were determined from different slices of 
the sky with absolutely no overlaps. This is a clinching evidence that 
some local clustering is not the cause of the inferred high velocity values.
The NVSS survey could be affected by a number of systematics, caused by changes in observing conditions when 
sweeping large areas on the sky, that may produce excess power on large angular scales. For example, Blake and Wall (2002) 
have noted that dim sources in NVSS show some declination dependence which largely disappears above 15 mJy, 
but nevertheless one cannot be totally sure that this and other possible systematics (e.g. calibration 
changes in time during the observing months) do not introduce a spurious modulation of the source density 
on large scales that may partially mimic a dipole. Again, except in a very contrived 
situation we could not have obtained consistent results for the 
projected velocity values. 

The expected CMBR values are way below the observed ones in Fig.~2, but one must 
ensure that there are no scaling problems either. 
Actually the plotted values are arrived at directly from the observed 
asymmetry in number counts in a straightforward calculation. For example, 
for the first plotted point ($\theta\sim32^{\circ}$) the observed number counts 
in the respective sky zones give a fractional number ${\Delta {\cal N}}/{{\cal N}}=0.0138$, which 
multiplied with $c/(2+x(1+\alpha))$ gives 1089 (km/s) as the projected 
velocity value, what is plotted in Fig.~2. Similar are the calculations for the other plotted 
points, and it does not seem that there could be anything amiss in the scaling. 
The evidence seems irrefutable that the velocity inferred from the radio 
source distribution is indeed much larger than that from the CMBR.
 
The fact that the directions of the dipole from the radio source data and the CMBR measurements are matching well, implies 
that the cause of the dipoles is common and the motion of the solar system seems to be the only reasonable explanation for 
that. But such a statistically significant difference in the estimates of the magnitude of the velocity vector is puzzling. 
Assuming that the CMBR dipole estimates do not suffer from any residual errors during subtraction of galactic and other 
contributions, one cannot escape the conclusion that there is a genuine discrepancy in the two dipoles and  
that the reference frame defined by the radio source population at  $z \sim 1$ does not coincide with that defined by the 
CMBR originating at $z \sim 700$. Here we may add that there is some evidence that the motion of the local group of 
galaxies may be different when measured with respect to different reference frames 
(Lauer and Postman 1994; Giovanelli et al. 1998) at 
$z\stackrel{<}{_{\sim}} 0.03$ and at $z\sim 0.05$. However, while an anisotropy at $z\stackrel{<}{_{\sim}} 0.05$ scale 
might still be called ``local'', an anisotropy at $z\stackrel{>}{_{\sim}} 1$ is nevertheless ``global'' as it encompasses 
a substantial fraction of the universe. 
On the other hand if there are  differential perturbations to the Hubble flow on the scale of the distribution of the 
radio source population vis-\'{a}-vis that 
of the CMBR, the implication will be serious as any such anomaly would imply anisotropy on a universal scale. 
This would violate the cosmological principle where the 
isotropy of the Universe is assumed for all epochs, and on which the whole modern cosmology rests upon. 

\end{document}